\documentclass[twocolumn,aps,superscriptaddress,showpacs]{revtex4}

\usepackage{amssymb}
\usepackage{amsmath}
\usepackage{graphicx}
\usepackage[normalem]{ulem}
\usepackage[dvips]{color}

\begin{document}
\title{Shear viscosity of neutron-rich nucleonic matter near its liquid-gas phase transition}

\author{Jun Xu}
\affiliation{Shanghai Institute of Applied
Physics, Chinese Academy of Sciences, Shanghai 201800, China}

\author{Lie-Wen Chen}
\affiliation{Department of Physics and
Astronomy and Shanghai Key Laboratory for Particle Physics and
Cosmology, Shanghai Jiao Tong University, Shanghai 200240, China}
\affiliation{Center of Theoretical Nuclear Physics, National
Laboratory of Heavy Ion Accelerator, Lanzhou 730000, China}

\author{Che Ming Ko}
\affiliation{Cyclotron Institute and
Department of Physics and Astronomy, Texas A$\&$M University,
College Station, Texas 77843, USA}

\author{Bao-An Li}
\affiliation{Department of Physics and
Astronomy, Texas A$\&$M University-Commerce, Commerce, TX
75429-3011, USA}
\affiliation{Department of Applied Physics, Xi'an
Jiao Tong University, Xi'an 710049, China}

\author{Yu-Gang Ma}
\email{ygma@sinap.ac.cn} \affiliation{Shanghai Institute of Applied
Physics, Chinese Academy of Sciences, Shanghai 201800, China}

\begin{abstract}
Within a relaxation time approach using free nucleon-nucleon cross
sections modified by the in-medium nucleon masses that are
determined from an isospin- and momentum-dependent effective
nucleon-nucleon interaction, we investigate the specific shear
viscosity ($\eta/s$) of neutron-rich nucleonic matter near its
liquid-gas phase transition. It is found that as the nucleonic
matter is heated at fixed pressure or compressed at fixed
temperature, its specific shear viscosity shows a valley shape in
the temperature or density dependence, with the minimum located at
the boundary of the phase transition. Moreover, the value of
$\eta/s$ drops suddenly at the first-order liquid-gas phase
transition temperature, reaching as low as $4\sim5$ times the KSS
bound of $\hbar/4\pi$. However, it varies smoothly for the
second-order liquid-gas phase transition. Effects of the isospin
degree of freedom and the nuclear symmetry energy on the value of
$\eta/s$ are also discussed.

\end{abstract}


\maketitle


Transport properties of hot nuclear matter at various densities,
such as the shear viscosity, can be extracted from model analyses of
heavy-ion collisions. In relativistic heavy-ion collisions, detailed
studies have shown that the produced Quark-Gluon-Plasma (QGP) has a
very small shear viscosity and behaves almost like an ideal
fluid~\cite{Gyu05,Shu05}. Specifically, it has been
found~\cite{Son11,Sch11} that the specific shear viscosity, i.e.,
the ratio of the shear viscosity to the entropy density, of the QGP
is only a few times the KSS lower bound of $\hbar/4\pi$ derived from
the AdS/CFT correspondence~\cite{Kov05}. Also, the specific shear
viscosity shows a minimum value around the critical temperature of
the hadron-quark phase transition~\cite{Cse06,Lac07}. It is argued
in Ref.~\cite{Cse06} that the existence of a minimum in the specific
shear viscosity is due to the difficulty for the momentum transport
in the QGP as its temperature is close to the critical temperature.

The shear viscosity of nucleonic matter is important for
understanding various phenomena, such as signatures of the possible
liquid-gas phase transition, in heavy-ion collisions at intermediate
energies~\cite{Dan84,Li11}. Because of the short-range repulsive and
intermediate-range attractive nature of the nucleon-nucleon
interaction, hot nucleonic matter is expected to undergo a
liquid-gas phase transition, see, e.g., Refs.~\cite{Fis67,Sie83}.
Imprints of such a phase transition on experimental observables,
such as the rank distribution of fragments~\cite{Ma}, are expected
in the multifragmentation process of heavy-ion collisions at
intermediate energies~\cite{Ber83}. However, while extensive studies
have been made to investigate both experimentally and theoretically
the signatures and nature of the liquid-gas phase transition using
various approaches and observables over the last thirty years, see,
e.g., Refs.~\cite{Gross01,Das05,Bor08} for recent reviews, many
interesting issues remain to be addressed. In fact, over the last
decade much work has been done to better understand the mechanism
and nature of the liquid-gas phase transition in isospin asymmetric
nucleonic matter, see, e.g., Refs.~ \cite{Ibook,WCI}. In particular,
what is the role of the isospin degree of freedom in nuclear
thermodynamics? What is the order of the liquid-gas phase transition
in neutron-rich nucleonic matter? What are the effects of the
density dependence of nuclear symmetry energy on the boundaries of
mechanical and chemical instabilities as well as the liquid-gas
coexistence line in neutron-rich matter? Answers to these questions
are important for understanding both astrophysical observations of
supernova explosions and terrestrial experiments done at rare
isotope beam facilities. However, many current answers are still
under debate. For instance, most models predict that while the
liquid-gas phase transition is of first order in isospin symmetric
matter, it becomes a continuous transition in isospin asymmetric
matter examined at a constant proton fraction. On the other hand, it
has been shown that the liquid-gas phase transition is actually
still of first order even in isospin asymmetric matter except at the
two ending points because of the existence of a spinodal
region~\cite{Duc06}.

Similar to its behavior at the hadron-quark phase transition, the
specific shear viscosity of nucleonic matter also shows a minimum
value at the vicinity of its liquid-gas phase
transition~\cite{Pal10,Zho12a,Zho12b}. Also, it was speculated that
the behavior of the specific shear viscosity at the phase transition
may depend on the order of the transition~\cite{Che07}. Thus,
further studies on the specific shear viscosity near the liquid-gas
phase transition may help shed new light on the nature of this
transition in neutron-rich matter. Indeed, it has been shown that
the boundaries of both mechanical and chemical instabilities
responsible for the phase separation~\cite{LiKo,Li01} and the phase
coexistence line~\cite{Xu07b,Xu08} in asymmetric nucleonic matter
depend on the value of the nuclear symmetry energy $E_{\rm
sym}(\rho)$ at subsaturation densities. It is, however, not known
how the $E_{\rm sym}(\rho)$ affects the specific shear viscosity of
nucleonic matter at the liquid-gas phase transition. Since the
nuclear matter can undergo the liquid-gas phase transition at
different temperatures and densities in intermediate-energy
heavy-ion collisions, it is of interest to know how the specific
shear viscosity would behave under these various conditions. For
example, is the valley shape structure in the temperature and
density dependence of the specific shear viscosity of nucleonic
matter the result of the liquid-gas phase transition?

In the present study, we use a relaxation time approach to study the
specific shear viscosity of neutron-rich nucleonic matter near the
liquid-gas phase transition based on a consistent Gibbs
construction. We find that the behavior of the specific shear
viscosity at the liquid-gas phase transition depends on its order,
and that the phase transition can cause a valley structure in the
temperature or density dependence of the specific shear viscosity,
although it does not necessarily require the existence of a phase
transition.


\begin{figure}[h]
\centerline{\includegraphics[scale=0.8]{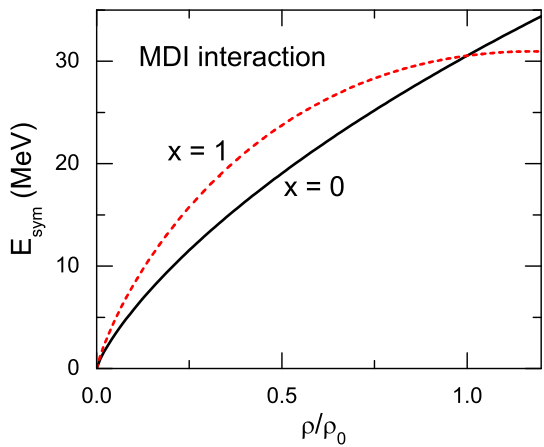}} \caption{(Color
online) Density dependence of a stiffer ($x=0$) and a
softer ($x=1$) symmetry energy from the MDI interaction.}
\label{Esym}
\end{figure}

For the nucleon-nucleon interaction, we use the isospin- and
momentum-dependent interaction proposed in Refs.~\cite{Das03,Che05}
(hereafter 'MDI') with its parameters fitted to the binding energy
$-16$ MeV and incompressibility $212$ MeV of normal nuclear matter
at the saturation density $\rho_0=0.16$ fm$^{-3}$. For the density
dependence of the symmetry energy, the parameter $x$ is used to
change its slope parameter $L=3\rho_0 \left(dE_{\rm
sym}/d\rho\right)_{\rho=\rho_0}$ but keeping its value at saturation
density fixed to $E_{\rm sym}(\rho_0)=31.6$ MeV. In particular, a
stiffer and a softer symmetry energy with $L\approx60$ MeV and
$L\approx15$ MeV are obtained with $x=0$ and $x=1$, respectively, as
shown in Fig.~\ref{Esym}, corresponding to current uncertainties in
the density dependence of the symmetry energy at subsaturation
densities~\cite{Che12}.

\begin{figure}[h]
\centerline{\includegraphics[scale=0.8]{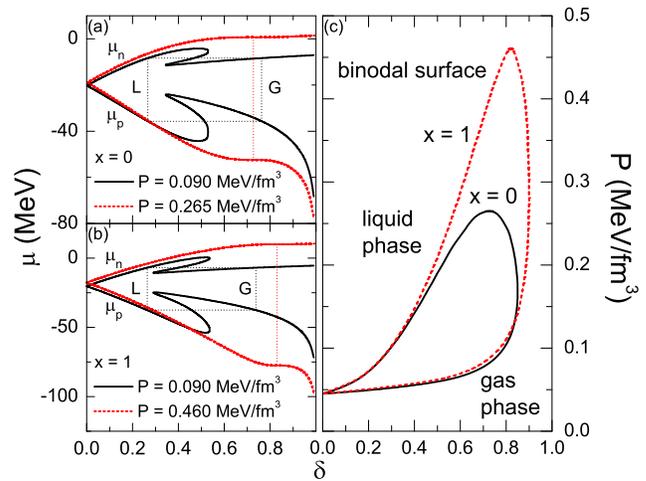}} \caption{(Color
online) Chemical potential isobar as a function of isospin asymmetry
for the stiffer ($x=0$) (a) and the softer ($x=1$) symmetry energies
(b) and binodal surface (c) for both values of $x$ at temperature
$T=10$ MeV.} \label{LGPT}
\end{figure}

To construct the liquid-gas phase transition region in the nuclear
phase diagram, we use the Gibbs conditions, i.e., the liquid and gas
phases can coexist when they have the same chemical potential
($\mu_l^{(n,p)}=\mu_g^{(n,p)}$), pressure ($P_l=P_g$), and
temperature ($T_l=T_g$). Specifically, we plot the chemical
potential isobar as a function of the isospin asymmetry $\delta$,
defined as $\delta=(\rho_n-\rho_p)/(\rho_n+\rho_p)$, for neutrons as
well as protons at a certain temperature, and draw a rectangle
within the proton and neutron chemical potential isobars. The two
ends of the rectangle then correspond to the two coexisting phases,
as shown in panels (a) and (b) of Fig.~\ref{LGPT} for the stiffer
($x=0$) and the softer ($x=1$) symmetry energies, respectively, with
the left end point having a smaller isospin asymmetry and a larger
density corresponding to a liquid phase (L) and the right end point
having a larger isospin asymmetry and a smaller density
corresponding to a gas phase (G). This procedure is repeated until
the pressure is too low to allow a rectangle to be drawn or too high
for the hot nucleonic matter to remain in the chemical instability
region, i.e., the chemical potential of neutrons (protons) increases
(decreases) monotonically with increasing isospin asymmetry. The
coexisting phases at different values of pressure form the binodal
surface shown in panel (c) of Fig.~\ref{LGPT}. The right and the
left side of the binodal surface correspond to the gas and the
liquid phase, respectively, with the mixed phase inside the binodal
surface. The binodal surface thus provides all the information
needed to study the properties of the mixed phase, i.e., the
densities and isospin asymmetries of the two coexisting phases as
well as their volume fractions. For more details on the liquid-gas
phase transition in nucleonic matter, we refer the readers to
Refs.~\cite{Mul95,Xu07b,Xu08}.

In the phase coexistence region with the liquid phase
occupying a volume fraction $\lambda$, the
average number and entropy densities can be expressed as
\begin{eqnarray}
\rho &=& \lambda\rho_l + (1-\lambda)\rho_g,\\
s &=& \lambda s_l + (1-\lambda)s_g,
\end{eqnarray}
where $\rho_{l(g)}$ and $s_{l(g)}$ are the number and entropy
densities of the liquid (gas) phase, respectively.

For the calculation of the shear viscosity, we consider a stationary
flow field in the z direction, i.e., $u_z=f(x)$ in the nucleonic
matter where $f(x)$ is an arbitrary function of the coordinate $x$,
and use a similar framework as in Ref.~\cite{Xu11}. For a single
phase of gas or liquid, the shear force on the particles in a flow
layer of a unit area in the $y-z$ plane is equal to the net
$z-$component of momentum transported per sec in the $x$ direction,
i.e., the thermal average of the product of the flux $\rho_\tau v_x$
in the $x$ direction and the momentum transfer $p_z-mu_z$ in the $z$
direction~\cite{Xu11,Hua87}
\begin{equation}
F_i = \sum_\tau \langle(p_z-mu_z)\rho_\tau v_x\rangle_i,
\end{equation}
with $\tau=n$ for neutrons and $p$ for protons, $i=l$ for the liquid
phase and $g$ for the gas phase, and $m$ being the nucleon mass. The
shear viscosity $\eta_{l(g)}$ is then determined by
\begin{equation}
F_{l(g)} = - \eta_{l(g)} \partial u_z/\partial x
\end{equation}
for either the liquid phase or the gas phase. We note that the shear
viscosity is independent of the flow gradient if $\partial
u_z/\partial x$ is sufficiently small.

For a mixed phase of liquid and gas, the matter can be viewed either
as gas bubbles in a liquid or liquid droplets in a gas. The matter
above and below any flow layer are then either both liquids or both
gas unless the flow layer is tangent to the surface of a gas bubble
or a liquid droplet, which would have the liquid and the gas on the
opposite sides of the flow layer. Since the chance for the latter to
happen is infinitesimally smaller for an infinitely large system
with liquid droplets and gas bubbles randomly distributed as assumed
in the present work, the fraction of the area for particle transport
across a flow layer in the liquid is thus $\lambda$ and that in the
gas is $1-\lambda$, leading to an average shear force on a unit area
of flow layer in the mixed phase given by the sum of the
contributions from individual phases, i.e.,
\begin{equation}
F = \lambda F_l + (1-\lambda) F_g = -\eta \partial u_z/\partial x.
\end{equation}
The average shear viscosity of the mixed phase can then be expressed
in terms of those in the liquid or the gas phase as
\begin{equation}
\eta = \lambda \eta_l + (1-\lambda) \eta_g.
\end{equation}
Because the density is uniform in each phase, $\eta_l$ and $\eta_g$
can be separately calculated using the relaxation time approach as
in Ref.~\cite{Xu11} based on free nucleon-nucleon cross
sections~\cite{Cha90} modified by the in-medium nucleon
masses~\cite{Li05}.

\begin{figure}[h]
\centerline{\includegraphics[scale=0.9]{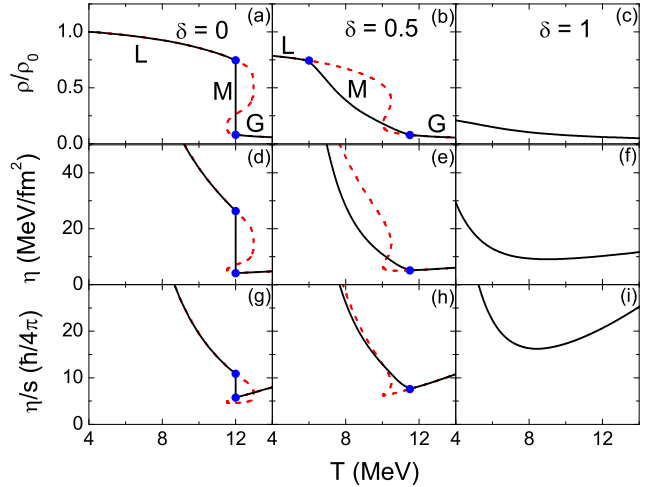}} \caption{(Color
online) Temperature dependence of the average reduced number density
(first row), the shear viscosity (second row), and the specific
shear viscosity (third row) at the fixed pressure of $P=0.1$
MeV/fm$^3$ for isospin symmetric matter ($\delta=0$) (left column),
neutron-rich matter ($\delta=0.5$) (middle column), and pure neutron
matter ($\delta=1$) (right column) with the stiffer symmetry energy
$x=0$. Solid lines are results including the liquid-gas phase
transition with 'L', 'M', and 'G' representing the liquid phase, the
mixed phase, and the gas phase, respectively. Dashed lines are
results obtained by assuming the liquid-gas phase transition does
not happen inside the binodal surface.} \label{etasT}
\end{figure}

Figure~\ref{etasT} displays the temperature dependence of the
average reduced number density, the shear viscosity, and the
specific shear viscosity, obtained with the stiffer symmetry energy
$x=0$, when the nucleonic matter is heated at the fixed pressure of
$P=0.1$ MeV/fm$^3$. As the temperature increases, the hot nucleonic
matter undergoes a phase transition from the liquid phase at lower
temperatures to the gas phase at higher temperatures if it has an
isospin asymmetry $\delta=0$ or $\delta=0.5$ but has no phase
transition if the isospin asymmetry is $\delta=1$. The liquid-gas
phase transition is of first order in symmetric nucleonic matter
($\delta=0) $ as shown in Fig.~26 of
Ref.~\cite{Xu08} by the sudden jump in the entropy per nucleon from
the liquid phase to the gas phase as well as the discontinuity of
the specific heat at the critical temperature. This leads to the
sudden changes in all the thermodynamical quantities and the
specific shear viscosity, while the latter evolves smoothly during
the phase transition when it changes to a second-order one in
neutron-rich matter ($\delta=0.5$), confirming the expectation of
Ref.~\cite{Che07}. Also, the liquid phase has a higher density and a
lower temperature than the gas phase as shown in the first row of
Fig.~\ref{etasT}, leading to a stronger Pauli blocking effect in the
liquid phase than in the gas phase. As a result, the liquid phase
generally has a larger shear viscosity than the gas phase. For each
phase, there are competing density and temperature effects on the
evolution of the shear viscosity. As discussed in Ref.~\cite{Xu11},
an increase in temperature results in more frequent nucleon-nucleon
scatterings and weaker Pauli blocking effects, thus reducing the
shear viscosity. On the other hand, the nucleon-nucleon scattering
cross section decreases with increasing center-of-mass energy of two
colliding nucleons as shown in Fig.~2 of Ref.~\cite{Xu11}, which
makes the shear viscosity to increase with increasing temperature
especially at very low densities. At higher densities, although the
stronger Pauli blocking effect increases the shear viscosity, the
smaller in-medium nucleon mass leads to a larger flux between flow
layers and a larger relative velocity between two colliding
nucleons, thus reducing the shear viscosity. Due to the combination
of these effects together with the behavior of the entropy density
with respect to temperature and density, the specific shear
viscosity decreases in the liquid phase but increases in the gas
phase with increasing temperature. The minimum of the specific shear
viscosity is exactly located at the critical temperature when a
first-order phase transition happens, while it is located at the
boundary of the gas phase if the phase transition is of second
order. Even for a pure neutron matter without a liquid-gas phase
transition, the specific shear viscosity still shows a valley shape
in its temperature dependence as a result of the competing effects
discussed above.

\begin{figure}[h]
\centerline{\includegraphics[scale=0.9]{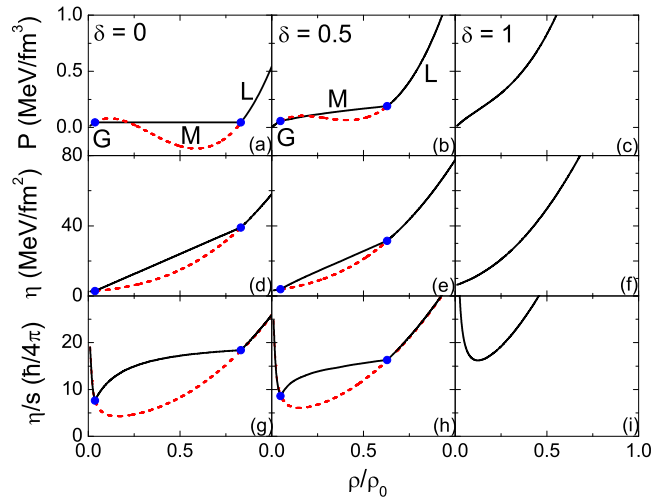}} \caption{(Color
online) Density dependence of the pressure (first row), the shear
viscosity (second row), and the specific shear viscosity (third row)
at temperature $T=10$ MeV for isospin symmetric matter ($\delta=0$)
(left column), neutron-rich matter ($\delta=0.5$) (middle column),
and pure neutron matter ($\delta=1$) (right column) with the stiffer
symmetry energy $x=0$. Solid lines are results including the
liquid-gas phase transition with 'L', 'M', and 'G' representing the
liquid phase, the mixed phase, and the gas phase, respectively.
Dashed lines are results obtained by assuming the liquid-gas phase
transition does not happen inside the binodal surface.}
\label{etasP}
\end{figure}

The liquid-gas phase transition can also happen if the hot nucleonic
matter is compressed at a fixed temperature. The density
dependence of the pressure, the shear viscosity, and the specific
shear viscosity in this case are shown in Fig.~\ref{etasP}, again
using the stiffer symmetry energy $x=0$. For the symmetric nuclear
matter ($\delta=0$) that has a first-order liquid-gas phase
transition, the pressure remains a constant when it is compressed
from the low-density gas phase to the high-density liquid phase. As
the nucleonic matter becomes neutron-rich ($\delta=0.5$) with the
phase transition changing to a second-order one, the pressure
continues to increase with increasing density in the mixed phase.
For the pure neutron matter, it again does not show a liquid-gas
phase transition when it is compressed at a fixed temperature.
It is shown in the second row of Fig.~\ref{etasP} that the
occurrence of the mixed phase in the hot nucleonic matter when it is
compressed at a constant temperature generally increases the
value of the shear viscosity compared with the case by assuming that
the liquid-gas phase transition does not happen. Also, the specific
shear viscosity always has a minimum value, and we found that this
is due to the difference in the increase of the shear viscosity and
the entropy density with increasing density, even for the case of
pure neutron matter without a liquid-gas phase transition.
Interestingly, the density at which the specific shear viscosity has
a minimum value is again located at the boundary of the gas phase
for $\delta=0$ and $\delta=0.5$ independent of the phase transition
order.

\begin{figure}[h]
\centerline{\includegraphics[scale=0.9]{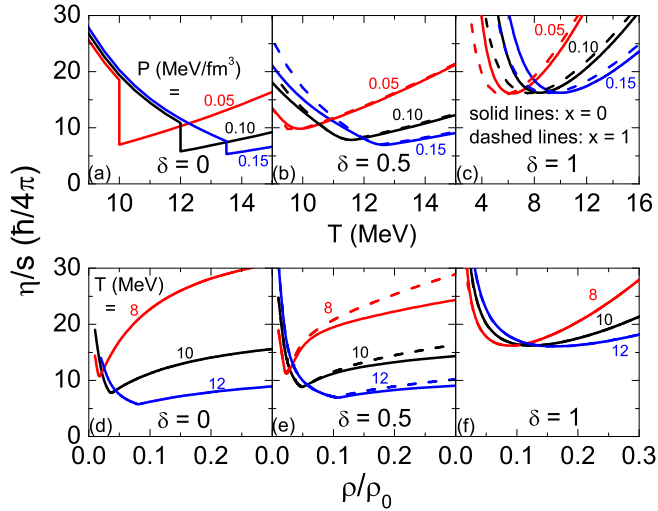}} \caption{(Color
online) Temperature (upper panels) and density (lower panels)
dependence of the specific shear viscosity at different fixed
pressures and temperatures, respectively, in isospin symmetric
matter ($\delta=0$), neutron-rich matter ($\delta=0.5$), and pure
neutron matter ($\delta=1$) for both symmetry energies $x=0$ and
$x=1$. } \label{etasPT}
\end{figure}

Since different values of pressure and temperature are reached in
intermediate-energy heavy-ion collisions, it is of interest to study
the specific shear viscosity of nucleonic matter at the liquid-gas
phase transition under different conditions. In the upper panels of
Fig.~\ref{etasPT}, we compare the temperature dependence of the
specific shear viscosity at different pressures for isospin
symmetric ($\delta=0$) and neutron-rich ($\delta=0.5$) nucleonic
matter as well as pure neutron matter ($\delta=1$) with both the
stiffer ($x=0$) and the softer ($x=1$) symmetry energies. It is seen
that the temperature at which the specific shear viscosity has a
minimum increases with increasing value of the fixed pressure,
similar to the results in Refs.~\cite{Cse06,Che07}. Also, for larger
fixed pressures the minimum value of the specific shear viscosity is
smaller for $\delta=0$ and $\delta=0.5$ but seems to be independent
of the pressure for $\delta=1$. In the lower panels of
Fig.~\ref{etasPT}, we display the density dependence of the specific
shear viscosity for different temperatures. Similarly, the density
at which the specific shear viscosity has a minimum value increases
with increasing value of the fixed temperature, and the minimum
value is smaller at higher fixed temperatures for $\delta=0$ and
$\delta=0.5$ but is insensitive to the temperature for $\delta=1$.
It is worthwhile to note that with further increase in pressure or
temperature, the minimum value of the specific shear viscosity
decreases and then levels off until the pressure or the temperature
is too high for the nucleonic matter to have a liquid-gas phase
transition. The resulting lower limit of the specific shear
viscosity of nucleonic matter is about $4\sim5$ $\hbar/4\pi$ for
isospin symmetric nucleonic matter and is generally smaller than
that in neutron-rich nucleonic matter as discussed in
Ref.~\cite{Xu11}. As seen in panel (c) of Fig.~\ref{LGPT}, the
stiffness of the symmetry energy only slightly affects the gas side
of the phase boundary, thus having only negligible effects on the
location of the minimum value of the specific shear viscosity.
However, due to the difference in the phase coexistence region for
the stiffer ($x=0$) and the softer ($x=1$) symmetry energy,
different specific shear viscosities are obtained in the mixed-phase
region, with the softer symmetry energy ($x=1$) giving a larger
value compared with the stiffer symmetry energy ($x=0$) as shown in
panels (b) and (e) of Fig.~\ref{etasPT}. For the pure neutron matter
without the liquid-gas phase transition under a fixed
pressure, it looks like that the specific shear viscosity for $x=1$
is similar to that for $x=0$ obtained under a slightly smaller fixed
pressure. On the other hand, the specific shear viscosities from
different symmetry energies are the same for pure neutron matter
compressed at a fixed temperature as indicated in panel (f) of
Fig.~\ref{etasPT}.

To summarize, using the relaxation time approach, we have studied
the specific shear viscosity of neutron-rich nucleonic matter near
its liquid-gas phase transition boundary constructed from the Gibbs
conditions. A valley shape is observed in the temperature or density
dependence of the specific shear viscosity even in the absence of
the phase transition. The value of the specific shear viscosity
suddenly drops at the first-order liquid-gas phase transition
temperature, while it varies smoothly for the second-order phase
transition.  Moreover, the density dependence of the symmetry energy
is found to affect the value of the specific shear viscosity of
nucleonic matter in the mixed-phase region, although it has little
effects on the location of its minimum. Our results are expected to
be useful for investigating the nature and signatures of the
liquid-gas phase transition in neutron-rich matter using
intermediate-energy heavy-ion collisions induced by rare isotopes.

\begin{acknowledgments}
We thank Yunpeng Liu for helpful discussions. This work was
supported by the "100-talent plan" of Shanghai Institute of Applied
Physics under grant Y290061011 from the Chinese Academy of Sciences,
the Major State Basic Research Development Program in China under Contract
No. 2014CB845401,
the NNSF of China (11135011, 11275125, 11035009, and 11220101005),
the Shanghai Rising-Star Program (11QH1401100), Shanghai "Shu Guang"
Project, the Eastern Scholar Program, and the STC of Shanghai
Municipality (11DZ2260700), the CUSTIPEN (China-U.S. Theory
Institute for Physics with Exotic Nuclei) under DOE grant number
DE-FG02-13ER42025, the US National Science Foundation under Grant
No. PHY-1068572 and PHY-1068022, the Welch Foundation under Grant
No. A-1358, and the National Aeronautics and Space Administration
under grant NNX11AC41G issued through the Science Mission
Directorate.
\end{acknowledgments}

\end{document}